\documentclass[aps,pre,onecolumn,superscriptaddress,showpacs]{revtex4-1}
\usepackage{graphicx}
\usepackage{dcolumn}
\usepackage{bm}

\usepackage{natbib}
\usepackage[
text={7in,9.5in},centering,
]{geometry}

\usepackage{float}
\usepackage{epsfig}
\usepackage{amsmath}
\usepackage{amssymb}
\usepackage{textcomp}
\usepackage{layouts}

\begin{document}

\title{Feynman Functional Integral in the Fokker Theory}

\author{Natalia Gorobey}
\affiliation{Peter the Great Saint Petersburg Polytechnic University, Polytekhnicheskaya
29, 195251, St. Petersburg, Russia}

\author{Alexander Lukyanenko}\email{alex.lukyan@mail.ru}
\affiliation{Peter the Great Saint Petersburg Polytechnic University, Polytekhnicheskaya
29, 195251, St. Petersburg, Russia}

\author{A. V. Goltsev}
\affiliation{Ioffe Physical- Technical Institute, Polytekhnicheskaya
26, 195251, St. Petersburg, Russia}

\begin{abstract}
The equivalence of two formulations of Fokker's quantum theory is proved - based on the Feynman functional integral representation of the propagator for a system of charges with direct electromagnetic interaction and the quantum principle of least action as an analogue of the Schr\"{o}dinger wave equation. The common basis for the two approaches is the generalized canonical form of Fokker's action.
\end{abstract}


\maketitle

\section{introduction}
Wheeler and Feynman \cite{1,2} formulated classical electrodynamics without the concept of electromagnetic field
in terms of the Fokker principle of least action \cite{3}. For
the simplest system of two particles, the Fokker action
has the form:
\begin{eqnarray}
 I=-m_1 c \int_0^{S_1}ds_1-m_2 c \int_0^{S_2}ds_2
 \notag\\
 +e_{1}e_2 \int_{0}^{S_1}ds_1 \int_{0}^{S_2}ds_2 \delta(s_{12}^2)\dot{x}_{1 \mu}{x'}_2^\mu,
 \label{1}
\end{eqnarray}
where
\begin{equation}
 s_{12}^2=(x_{1 \mu}(s_1)-x_{2 \mu}(s_2))(x_1^\mu(s_1)-x_2^\mu(s_2)).
\end{equation}
The dot denotes the derivative with respect to $s_1$, and the prime denotes the derivative with respect to $s_2$. Despite the multi-temporal nature of dynamics in the Fokker theory, it admits a generalized canonical formulation in which the world lines of
particles as a whole serve as canonical coordinates \cite{4,5}. This formulation is proposed in \cite{6}. Within this formulation, the generalized canonical momenta are defined as variational derivatives of action Eq.(\ref{1}) with respect to velocities:
\begin{eqnarray}
  p_{1 \mu}(s_1)=\frac{\delta{I}}{\delta{\dot{x}_1^\mu(s_1)}}=-m_1 c \dot{x}_{1\mu}(s_1)
 \notag\\
 +e_{1} e_2\int_0^{S_2}ds_2\delta(s_{12}^2)x'_{2 \mu}(s_2),
 \label{3}
\end{eqnarray}
\begin{eqnarray}
 p_{2 \nu}(s_2) = \frac{\delta I}{\delta {x'}_2^\nu (s_2)}=-m_2 c{x'}_{2 \nu}(s_2)
 \notag\\
 +e_{1} e_{2} \int_{0}^{S_1} d s_{1} \delta(s_{12}^2) \dot{x}_{1 \nu}(s_1),
 \label{4}
\end{eqnarray}
and the Hamilton function (functional) is found using the generalized Legendre transform:
\begin{eqnarray}
 H[x_1,p_1;x_2,p_2]
 \notag\\
 =\int_0^{S_1}ds_1 p_{1 \mu} \dot{x}_1^\mu+\int_0^{S_2} ds_2 p_{2\nu}{x'}_2^\nu
 \notag\\
 -I[x_1,\dot{x}_1;x_2,x'_2].
\end{eqnarray}
Here, as usual, velocities are eliminated by solving equations Eq.(\ref{3}) and Eq.(\ref{4}). In this case, this solution can be represented by a series of perturbation theory on $e_1e_2$ and we obtain the Hamilton functional in the form
\begin{equation}
H[x_1,p_1;x_2,p_2]=\textbf{M}-m_1^2 c^2 S_1-m_2^2 c^2 S_2
\end{equation}
\begin{eqnarray}
 \textbf{M} = \sum\limits_{\alpha,\beta=1,2} \int_0^{S_\alpha} ds_\alpha \int_0^{S_\beta} ds_\beta M_{\alpha \mu \beta \nu}(x_\alpha,x_\beta)
\notag\\
\times p_{\alpha \mu}(s_\alpha) p_{\beta \nu} (s_\beta).
\end{eqnarray}
We do not need the explicit form $\textbf{M}$ here. After that, we obtain the generalized canonical form of Fokker's action:
\begin{eqnarray}
 I_H[x_1,p_1;x_2,p_2]
 \notag\\
 =\int_0^{S_1}ds_1p_{1 \mu}\dot{x}_1^\mu+\int_0^{S_2}ds_2p_{2 \nu}{x'}_2^\nu
 \notag\\
 -H[x_1,p_1;x_2,p_2].
 \label{8}
\end{eqnarray}

All this was necessary for us in order to define the propagator of this system in quantum theory in the form of a functional integral on the generalized phase space. By analogy with ordinary quantum mechanics, where this functional integral is obtained as a representation of the kernel of the evolution operator for the Schr$\ddot{o}$dinger equation \cite{7}, we can write \cite{6}:
\begin{eqnarray}
 K(\tilde{x}_1,S_1,\tilde{x}_2,S_2;x_1,0,x_2,0)
 \notag\\
=\int\frac{d^4 \tilde{p}_1}{(2\pi{\hbar})^4}\prod_{s_1,\mu}\frac{dp_{1 \mu} dx_{1 \mu}}{2\pi{\hbar}} \frac{d\tilde{p}_2}{2\pi{\hbar}}\prod_{s_2,\nu}\frac{dp_{2 \nu}dx_{2 \nu}}{2\pi{\hbar}}
\notag\\
\times{\exp\left\{\frac{i}{\hbar}I_H[x_1,p_1;x_2,p_2]\right\}}
\label{9}
\end{eqnarray}
In this paper, the quantum principle of least action is proposed as a dynamic justification for this functional integral. This principle serves as a generalization of the Schr\"{o}dinger equation for the case of a multi-time dynamical system.

\section{Quantum principle of least action}

First, the quantum principle of least action was formulated as a new form of local quantum mechanics with one parameter of time \cite{8}, and then generalized to the case of multi-time dynamical theory \cite{9}. In this formulation of quantum mechanics, the generalized canonical momenta are replaced by the operators of variational differentiation,
\begin{equation}
 \hat{p}_{1 \mu}(s_1)=\frac{\tilde{\hbar}}{i}\frac{\delta}{\delta{x}_1^\mu(s_1)},
 \label{10}
 \end{equation}
\begin{equation}
 \hat{p}_{2 \nu}(s_2)=\frac{\tilde{\hbar}}{i}\frac{\delta}{\delta{x}_2^\nu(s_2)}
 \label{11}
\end{equation}
acting in the space of wave functionals, which are defined in the space of world lines of particles, $\Psi[x_1(s_1),x_2(s_2)] (\tilde{x}_1,S_1;\tilde{x}_2,S_2)$. These functionals are at the same time functions of the boundary points of the world lines; we have indicated this dependence on the end points explicitly in parentheses. We will need this explicit dependence in the future.
The generalized Planck's constant also appears here,

\begin{equation}
 \tilde{\hbar}=\hbar{\sigma}.
\end{equation}
The factor $\sigma$ of the dimension of proper time will be defined below. Let us substitute the operators Eqs. (\ref{10}) and (\ref{11}) in the generalized canonical action Eq. (\ref{8}) placing them on the right in all terms. We obtain the operator of action on the space of wave functionals. The quantum principle of least action is a problem for the eigenfunctions and eigenvalues of the action operator:
\begin{equation}
 \hat{I}_H\Psi=\Lambda{\Psi}.
 \end{equation}
The eigenvalue is a function of the start and end points of the particle world lines only: $\Lambda=\Lambda(\tilde{x}_1,S_1;\tilde{x}_2,S_2)$.

Next, you should keep in mind the finite-dimensional representation of the formalism, replacing the world lines of the particles with broken lines. To do this, we divide the proper time intervals $[0,S_1],[0,S_2]$ into small sections of length $\epsilon_1,\epsilon_2$, and denote the vertices of the broken lines by $x_{1 n}^\mu,x_{2 m}^\nu$. The wave functional is then represented as a function of many variables - the vertices of the broken lines,
\begin{equation}
 \Psi=\Psi(x_{1 n},x_{2 m}),
\end{equation}
where the end points are $x_{1 N}=\tilde{x}_1,x_{2 M}=\tilde{x}_2$. We keep the previous notation of the wave functional.

Now for the constant $\sigma$. It is a measure of the nonlocality of particle interaction at a finite speed of signal propagation,
\begin{equation}
 \sigma=\frac{D}{c}
\end{equation}
where $D$ is a quantity proportional to the size of the particle interaction region, which is determined by the experimental conditions, say, in the scattering problem. We can fix this quantity from the principle of correspondence of the nonlocal quantum theory considered here with local quantum mechanics in the nonrelativistic limit $c\rightarrow\infty$.
In this case, $\sigma\rightarrow0$, and we obtain a singular wave functional in the form of the product of the values of the wave function at different moments of the total time $t$ of the particles:
\begin{equation}
 \Psi=\prod_{n}\psi(t_n,x_{1 n},x_{2n})
\end{equation}
where $t_n =\sigma n$. The singularity here also manifests itself in the fact that the second variational derivatives,
\begin{equation}
 \frac{{\delta}^2\Psi}{\delta{x}^2(s)}\propto\sigma^{-1},
\end{equation}
in the action operator in the nonrelativistic limit turn to infinity.

\section{Dynamic justification of the functional integral}

Let us see where the quantum principle of least action leads as applied to the last small regions $[S_1-\epsilon_1,S_1], [S_2-\epsilon_2,S_2]$
particle motion. The part of the Hamilton functional corresponding to this stage of particle motion has the following form:
\begin{equation}
 \delta{H}=\delta{\textbf{M}}-m_1^2c^2\epsilon_1-m_2^2c^2\epsilon_2
\end{equation}
where
\begin{eqnarray}
\delta \textbf{M}=\sum\limits_{\alpha,\beta=1,2}\int_0^{S_\beta}ds_\beta \epsilon_\alpha{M_{\alpha \mu \beta \nu}(x_\alpha,x_\beta)}
\notag\\
\times{p_{\alpha \mu}(s_\alpha)p_{\beta \nu}(s_\beta)}
\notag\\
+\sum\limits_{\alpha,\beta=1,2} \int_0^{S_\alpha}ds_\alpha \epsilon_\beta M_{\alpha \mu \beta \nu}(x_\alpha,x_\beta)
\notag\\
\times p_{\alpha \mu}(s_\alpha) p_{\beta \nu}(s_\beta).
\end{eqnarray}
The quantum principle of least action itself can be written as
\begin{equation}
\delta \hat{I}_H\Psi=\delta \Lambda \Psi.
\label{20}
\end{equation}
One of the tools for obtaining a representation of the kernel of the evolution operator in the form of a functional integral in \cite{7} is the following representation of the $\delta$-function:
\begin{equation}
 \delta^{(4)}(x-y)=\int\frac{d^4p}{(2\pi \hbar)^4}\exp[\frac{i}{\hbar}p(x-y)],
\end{equation}
where we use the condensed notation for the dot product in the Minkowski space. We also need its functional generalization:
\begin{eqnarray}
\prod_{s} \delta^{(4)}(x(s)-y(s))=\int\prod_{s} \frac{d^4p(s)}{(2\pi \tilde{\hbar})^4}
\notag\\
\times \exp[\frac{i}{\tilde{\hbar}}\int_0^{S} ds p(s)(x(s)-y(s))].
\end{eqnarray}
Using this tool, the action of the generalized momentum operators Eqs. (\ref{10}) and (\ref{11}) is reduced to multiplication by the momenta (at the very end of formula below), for example,
\begin{eqnarray}
 {\hat{p}_{1 \mu}(s_1)}\Psi[x_1(s_1),x_2(s_2)]
 \notag\\
=\int\prod_{s_1}\frac{d^4 p_1(s_1)}{2 \pi \tilde{\hbar}} \prod_{s_2}\frac{d^4 p_2(s_2)}{2 \pi \tilde{\hbar}}
\notag\\
\times{\prod_{s_1}}d^4y_1(s_1)\prod_{s_2} d^4y_2(s_2)
\notag\\
\times \exp[\frac{i}{\tilde{\hbar}}\int_0^{S_1}ds_1 p_1(s_1)(x_1(s_1)-y_1(s_1))]
\notag\\
\times\exp[\frac{i}{\tilde{\hbar}}\int_0^{S_2}ds_2 p_2(s_2) (x_2(s_2)-y_2(s_2))]
\notag\\
\times p_{1 \mu}(s_1)\Psi[y_1(s_1),y_2(s_2)]
\end{eqnarray}
But for the very last stage $[S_1-\epsilon_1,S_1],[0,S_2]$ of the dynamics of the first particle, the same result can be obtained using the usual $\delta$-function:
\begin{eqnarray}
 \frac{\hbar}{i}\frac{\partial}{\partial \tilde{x}_1^\mu}\Psi[x_1(s_1),x_2(s_2)]
 \notag\\
=\int \frac{d^4 \tilde{p}_1}{(2\pi \hbar)^4}\prod_{s_2} \frac{d^4 p_2(s_2)}{(2 \pi \tilde{\hbar})^4}
\notag\\
\times d^4y_{1 N{-}1}\prod_{s_2} d^4y_2(s_2)
\notag\\
\times \exp[\frac{i}{\hbar}\tilde{p}_1(\tilde{x}_1-y_{1 N{-}1})]
\notag\\
\times \exp[\frac{i}{\tilde{\hbar}} \int_0^{S_2} ds_2 p_2(s_2)(x_2(s_2)-y_2(s_2))]
\notag\\
\times \tilde{p}_{1 \mu}\Psi[y_1(s_1),y_2(s_2)].
\end{eqnarray}
The same is true for the second particle. Similarly, the first term in the action operator for the first particle in Eq. (\ref{20}) can be written as
\begin{eqnarray}
 (\tilde{x}_1 - y_{1 N{-}1}) \frac{\hbar}{i}\frac{\partial}{\partial \tilde{x}_1}\Psi[x_1(s_1),x_2(s_2)]\sim
 \notag\\
-(\tilde{x}_1-y_{1 N{-}1})\frac{\hbar}{i}\frac{\partial}{\partial y_{1 N{-}1}}\Psi[y_1(s_1),y_2(s_2)]
\notag\\
=-\epsilon_1 \frac{(\tilde{x}_{1}-y_{1 N{-}1})}{ \epsilon_1} \frac{\hbar}{i} \frac{\partial}{\partial y_{1 N{-}1}}\Psi[y_1(s_1),y_2(s_2)]
\notag\\
=-\frac{\hbar}{i}{\epsilon_1}\frac{\partial}{\partial S_1}\Psi[y_1(s_1),y_2(s_2)]
\notag\\
+\delta_1{\Lambda}\Psi[y_1(s_1),y_2(s_2)],
\end{eqnarray}
where the tilde sign denotes all the omitted operations of the previous formula. The corresponding contribution to the eigenvalue of the action operator is:
\begin{equation}
 \delta_1{\Lambda}=\frac{\hbar}{i}[\ln\Psi(\tilde{x}_1,S_1;\tilde{x}_2,S_2)-\ln\Psi(\tilde{x}_1,S_1-\epsilon_1;\tilde{x}_2,S_2)]
\end{equation}
We get exactly the same contributions for the second particle.
Summing up both contributions and repeating this reasoning many times, we completely replace the functional delta function with the usual one at all stages of quantum dynamics of particles. The result is the required functional integral on the generalized phase space \cite{6}. This integral, in turn, defines the eigenvalue of the action operator by the following formula:
\begin{equation}
\Lambda=\frac{\hbar}{i} \ln K(\tilde{x}_1,S_1,\tilde{x}_2,S_2;x_1,0,x_2,0)
\end{equation}
Thus, the justification of formula Eq. (\ref{9}) using the quantum principle of least action is completed. The opposite can be said: Feynman's functional integral substantiates the quantum principle of least action. Both formulations of Fokker's quantum theory are equivalent.

\section{Conclusions}
The equivalence of different formulations allows us to consider the Fokker quantum theory as a reliable basis for the theory of real processes of electromagnetic interactions without introducing the concept of an electromagnetic field and its quantization. This was the original vision of Wheeler and Feynman. They implemented it in the classical theory, improving the Fokker theory with the concept of an absorber. For quantum theory, Feynman developed the functional integral formalism. It can be expected that the effects of quantum electrodynamics will also be taken into account in the quantum theory of Fokker, if the effect of all charges in the universe on the system under consideration is properly taken into account.
Using the proper time of each charge as an independent evolution parameter allows us to include the processes of pair production and annihilation into consideration.
However, for comparison with experiment, these parameters should be fixed by the additional principle of the extremum of the quantum action. The formalism proposed in this work allows us to do this.

\section{Acknowledgement}

The authors thank V.A. Franke for useful discussions.


\begin{thebibliography}{99}

\bibitem{1} R. P. Feynman and J. A. Wheeler, Rev. Mod. Phys. \textbf{17}, 157 (1945).

\bibitem{2} R. P. Feynman and J. A. Wheeler, Rev. Mod. Phys. \textbf{21}, 425 (1949).

\bibitem{3} A. D. Fokker, Z. Phys. \textbf{58}, 386 (1929).

\bibitem{4} X. Jaen, R. Jauregui, J. Llosa, and A. Molina, Phys. Rev. D \textbf{36}, 2385 (1987).

\bibitem{5} X. Jaen, R. Jauregui, J. Llosa, and A. Molina, J. Math. Phys. \textbf{30}, 2807-2814 (1989).

\bibitem{6} Natalia Gorobey, Alexander Lukyanenko, and A. V. Goltsev, arXiv:2002.03607v1 [quant-ph] (2020).

\bibitem{7} L. D. Faddeev and A. A. Slavnov, Gauge fields: An introduction to quantum theory, Westview Press, 2d edition, 236 (1993).

\bibitem{8} N. N. Gorobey and A. S. Lukyanenko, arXiv:0807.3508v1 [quant-ph] (2008).

\bibitem{9} Natalia Gorobey, Alexander Lukyanenko, and Inna Lukyanenko, arXiv:0910.2157v1 [quant-ph] (2009).

\end{thebibliography}
\end{document}